\documentclass[smallextended,referee,envcountsect]{svjour3}
\smartqed
\usepackage{graphicx,amssymb,latexsym,amsmath,enumerate,verbatim,amsfonts}
\journalname{JOTA}

\begin{document}


\title{Copositivity for a class of fourth order symmetric tensors given by scalar dark matter}
\subtitle{ }

\author{Yisheng Song   \and  Xudong Li }

\institute{Yisheng Song \ Corresponding author \at
              School of Mathematical Sciences, Chongqing Normal University, Chongqing, P.R. China, 401331
                \email{yisheng.song@cqnu.edu.cn}
           \and
           Xudong Li   \at
             School of Mathematics and Information Science,
        Henan Normal University, XinXiang HeNan,  P.R. China, 453007
               \email{1516512738@qq.com}
            }

\date{Received: date / Accepted: date}

\maketitle

\begin{abstract}
The mathematical model of multiple microscopic particles potentials corresponds to a fourth order symmetric tensor with a particular structure in particle physics. In this paper, we mainly dedicate to the study of copositivity for a class of tensors defined by the scalar dark matter with the standard model Higgs  and an inert doublet and a complex singlet. With the help of its structure, we obtain the necessary and sufficient conditions, which attains the analytic conditions required by the physical problems.  At the same time, this analytic expression provides how to determine a unique solution of the corresponding tensor complementarity problem with a parameter.
\end{abstract}

\section{Introduction}

 In particle physics, the standard model of multiple real scalar fields or multiple microscopic particles potentials is a quartic homogeneous polynomial. For example, the most general scalar  potential of the scalar dark matter stable with the standard model Higgs, an inert doublet and a complex singlet is a quartic homogeneous polynomial (\cite{BK2012,BK2014,C15,C16,C17}). It is well-known that there is a one-to-one correspondence between a quartic homogeneous polynomial and a fourth-order symmetric tensor,  and then the mathematical model of the above scalar dark matter stable may be written as a fourth order three-dimensional symmetric tensor  with a parameter. So,  the vacuum stability of such a scalar dark matter is really equivalent to the  copositivity of the corresponding tensor.

 Qi \cite{C23,C24} first proposed the concept of copositivity and positive definiteness of symmetric tensor, and  showed  that a symmetric tensor is (strictly) copositive if the sum of diagonal element of each row and the all less than zero elements of the row is (positive) nonnegative. Subsequently, many good properties are obtained for this class of tensors.  By means of the properties of the principal sub-tensors, Song-Qi \cite{C28} proposed a new method to examine the (strict) copositivity of symmetric tensors. Song-Qi \cite{C33} introduced the notions of Pareto H-eigenvalue and Pareto Z-eigenvalue by the aid of Lagrange multipliers, and established the relations between the Pareto H-eigenvalue (Z-eigenvalue) and the (strict) copositivity of corresponding tensor.  Song-Qi \cite{C31} presented that the (strict) copositivity of a symmetric tensor is equivalent to its (strict) semi-positivity.  The notion of semi-positivity is firstly used by Song-Qi \cite{C29}, and this class of tensors have many nice properties and applications in the tensor complementary problems. Song-Qi \cite{C29} showed that every strictly semi-positive tensor is a Q-tensor. Moreover, we may take advantage of these conclusions to explore the copositivity of tensors and its applications \cite{C30,C34,C36,C38}. More details about tensor complementary problems  may be found in Refs.\cite{C1,C2,C3,C4,C8,C9,C10,C14,C20,C39,C40,C41} and others.

For testing copositivity of tensors, the numerical algorithms have been widely studied. Chen-Huang-Qi \cite{C5} attained several fundamental theories of copositivity of symmetric tensor and constructed corresponding numerical algorithms. Li-Zhang-Huang-Qi \cite{C18} gave a SDP relaxation algorithm of the copositivity of higher order tensors. Chen-Huang-Qi \cite{C6} and Chen-Wang \cite{C7} proposed several numerical algorithms to check the copositivity of high order tensors. For more details about copositivity algorithms, also see \cite{C21,C25,C27}. However, these conclusions and algorithms are not  specially designed for the fourth order symmetric tensors, and may not attain the analytic conditions required by the physical problems.

For the fourth order symmetric tensors, Song-Qi \cite{C32} and Liu-Song \cite{C19} presented several sufficient conditions of its copositivity; Song \cite{C37} proposed several sufficient conditions of its positive definiteness. Without considering the discriminant, Guo \cite{C11} provided the necessary and sufficient conditions of positive definiteness of a binary quartic form. Recently, Qi-Song-Zhang \cite{C26} obtained new necessary and sufficient conditions of the non-negativity  of a quartic polynomial with one variable. For a fourth order symmetric tensor defined in particle physics, Song-Qi \cite{C35} showed the necessary and sufficient conditions of its positive definiteness. But the analytic necessary and sufficient conditions of copositivity have not been obtained.

In this paper, we work on seeking analytic and checkable  conditions for copositivity of fourth order symmetric tensor. Motivated by Qi-Song-Zhang's  result \cite{C26}, we  propose a simple analytic expression of copositivity of fourth order two-dimensional symmetric tensor. With the assistance of these consequences, we give the analytic necessary and sufficient conditions of copositivity for a fourth order three-dimensional symmetric tensor given by the vacuum stability of scalar dark matter. Furthermore, this actually presents a method of identifying copositivity of the tensor  with a parameter, and then, this work also gives a method to test the uniqueness of solution to tensor complementary problems.

    \section{Preliminaries and Basic facts}

     \ \ \ \ An $m$th order $n$-dimensional symmetric tensor $\Upsilon=(\tau_{i_{1}{i_{2}}\ldots{i_{m}}})$ is said to be
\begin{flushleft}  (i) {\it (strictly) copositive} if for all non-negative vector $x$ with $\|x\|=1$, \begin{equation}\label{2}\Upsilon x^{m}=\sum\limits_{i_1,\cdots,i_m=1}^{n}\tau_{i_1i_2\cdots i_m}x_{i_1}x_{i_2}\cdots x_{i_m}\geq0\ (>0);\end{equation} \end{flushleft}
\begin{flushleft}
     (ii) {\it semipositive (positive) definite } if $\Upsilon x^{m}\geq0 \ (>0)$ for every vector $x \in\mathbb{R}^{n}$ with $\|x\|=1$ and an even number $m$;
     \end{flushleft}
     (iii) {\it  (strictly) semi-positive} if for each non-negative and non-zero vector $x$, there exists an index $i\in\{1,2,\ldots,n\}$ such that
     $$
     x_i>0\mbox{ and }(\Upsilon x^{m-1})_i=\sum\limits_{i_2,\cdots,i_m=1}^{n}\tau_{ii_2\cdots i_m}x_{i_2}\cdots x_{i_m}\geq0\ (>0).
     $$

For low order matrices, the copositive conditions are relatively simple. This gives the following Lemma about the copositivity of matrices. For more details  about these conclusions, see  Andersson-Chang-Elfving \cite{ACE}, Chang-Sederberg \cite{CS1994}, Hadeler \cite{H1983} and Nadler \cite{N1992}.

   \begin{lemma}\label{lem21}
   A symmetric   $2\times 2$ matrix $M=(m_{ij})$ is (strictly) copositive if and only if
   $$
   m_{11}\geq0\ (>0), m_{22}\geq0\ (>0), m_{12}+\sqrt{m_{11}m_{22}}\geq0\ (>0).
  $$
   A symmetric $3\times 3$ matrix $M=(m_{ij})$ is (strictly) copositive  if and only if
$$\begin{aligned}
m_{11}\geq 0\ (>0), m_{22}\geq 0\ (>0), m_{33}\geq 0\ (>0),
 \alpha=m_{12}+\sqrt{m_{11}m_{22}}\geq 0\ (>0),\\ \beta=m_{13}+\sqrt{m_{11}m_{33}}\geq 0\ (>0), \gamma=m_{23}+\sqrt{m_{33}m_{22}}\geq 0\ (>0),\\
 m_{12}\sqrt{m_{33}}+m_{13}\sqrt{m_{22}}+m_{23}\sqrt{m_{11}}+\sqrt{m_{11}m_{22}m_{33}}+\sqrt{2\alpha\beta\gamma}\geq 0\ (>0).
\end{aligned}$$
   \end{lemma}

 The conditions of a quadratic polynomial with one variable more than or equal to zero are well-known (also see Qi-Song-Zhang \cite{C26} for more details).
 
\begin{lemma}\label{lem23}
Let g(t) be a quadratic polynomial with one variable and $a>0$,
$$
g(t)=at^{2}+bt+c.
$$
 Then $s(t)>0\ (\geq0)$ for all $t\geq0$ if and only if
 $$\begin{cases} c>0\ (\geq0),\ &\mbox{ if }b\geq0,\\
4ac-b^{2}>0\ (\geq0),\ &\mbox{ if }b<0.
\end{cases}$$
\end{lemma}

The non-negativity of a quartic  polynomial with one variable is showed by Ulrich-Watson \cite{C13} for $s>0$. Recently, Qi-Song-Zhang \cite{C26} reexpressed their conclusions.

\begin{lemma}\label{lem24}
Let $f(s)$ be a quartic polynomial with $\alpha>0$ and $\eta>0$,
\begin{equation*}
f(s)=\alpha s^{4}+\beta s^{3}+\gamma s^{2}+\mu s+\eta.
\end{equation*}
Then  $s(t)$ is non-negative for all $s>0$ if and only if
\begin{flushleft}
\quad(1) $\Delta\leq0$ and $\beta\sqrt{\eta}+\mu\sqrt{\alpha}>0;$ or
\end{flushleft}
\begin{flushleft}
\quad(2) $\beta\geq0$, $\mu\geq0$ and $2\sqrt{\alpha\eta}+\gamma\geq0;$ or
\end{flushleft}
\begin{flushleft}
\quad(3) $\Delta\geq0$, $|\beta\sqrt{\eta}-\mu\sqrt{\alpha}|\leq4\sqrt{\alpha\gamma\eta+2\alpha\eta\sqrt{\alpha\eta}}$ and either
\end{flushleft}
\begin{flushleft}
\qquad(i) $-2\sqrt{\alpha\eta}\leq \gamma \leq6\sqrt{\alpha\eta}$, or
\end{flushleft}
\begin{flushleft}
\qquad(ii) $\gamma>6\sqrt{\alpha\eta}$ and $\beta\sqrt{\eta}+\mu\sqrt{\alpha} \geq -4\sqrt{\alpha\gamma\eta-2\alpha\eta\sqrt{\alpha\eta}}$,
\end{flushleft}
where $\Delta=4(12\alpha\eta-3\beta\mu+\gamma^{2})^{3}-(72\alpha\gamma\eta+9\beta\gamma\mu-2\gamma^{3}-27\alpha\mu^{2}-27\beta^{2}\eta)^{2}$.
\end{lemma}

\section{Main Results}
 Let $T=(t_{ijkl})$  be a 4th order 2-dimensional symmetric tensor. For any vector $x=(x_{1},x_{2})^\top$,
\begin{align*} \begin{split}
T x^{4}&=\sum_{i,j,k,l=1}^{2}t_{ijkl}x_{i}x_{j}x_{k}x_{l}\\
&=t_{1111}x_{1}^{4}+4t_{1112}x_{1}^{3}x_{2}+6t_{1122}x_{1}^{2}x_{2}^{2}+4t_{1222}x_{1}x_{2}^{3}+t_{2222}x_{2}^{4}.
\end{split} \end{align*}
Next, we show a analytical expression of the copositivity of tensor $\Upsilon$.

\begin{theorem}\label{T31}
 Let $T=(t_{ijkl})$a 4th order 2-dimensional symmetric tensor with $t_{1111}>0$ and $t_{2222}>0$. Then $\Upsilon$ is copositive if and only if
\begin{itemize}
  \item[(1)] $\Delta\leq0$, $t_{1222}\sqrt{t_{1111}}+t_{1112}\sqrt{t_{2222}}>0$; or
  \item[(2)] $t_{1222}\geq0$, $t_{1112}\geq0$, $3t_{1122}+\sqrt{t_{1111}t_{2222}}\geq0$; or
  \item[(3)] $\Delta\geq0$,\\
  \quad
  $|t_{1112}\sqrt{t_{2222}}-t_{1222}\sqrt{t_{1111}}|\leq\sqrt{6t_{1111}t_{1122}t_{2222}+2t_{1111}t_{2222}\sqrt{t_{1111}t_{2222}}}$

  (i) $-\sqrt{t_{1111}t_{2222}}\leq3t_{1122}\leq3\sqrt{t_{1111}t_{2222}}$;

(ii) $t_{1122}>\sqrt{t_{1111}t_{2222}}$,

 $t_{1112}\sqrt{t_{2222}}+t_{1222}\sqrt{t_{1111}}\geq-\sqrt{6t_{1111}t_{1122}t_{2222}-2t_{1111}t_{2222}\sqrt{t_{1111}t_{2222}}}$,
\end{itemize}
where \begin{align*}
\Delta &=4\times12^{3}(t_{1111}t_{2222}-4t_{1112}t_{1122}+3t_{1222}^{2})^{3}\\
&\quad-72^{2}\times6^{2}(t_{1111}t_{1122}t_{2222}+2t_{1112}t_{1122}t_{1222}-t_{1122}^{3}-t_{1112}^{2}t_{2222}-t_{1111}t_{1122}^{2})^{2}.
\end{align*}
\end{theorem}
\begin{proof} For $x=(x_{1},x_{2})^\top$ with $\|x\|=1$, and $x_{1}$, $x_{2}$ are non-negative, we have
$$T x^{4}=t_{1111}x_{1}^{4}+4t_{1112}x_{1}^{3}x_{2}+6t_{1122}x_{1}^{2}x_{2}^{2}+4t_{1222}x_{1}x_{2}^{3}+t_{2222}x_{2}^{4}.$$
It is obvious that $T x^{4}=t_{2222}>0$ if $x_{1}=0$ and $x_{2}\neq0$, and $T x^{4}=t_{1111}>0$ if $x_{2}=0$ and $x_{1}\ne0$.
Suppose $x_{1}>0$ and $x_{2}>0$, we may rewritten the homogeneous polynomial $Tx^{4}$,
$$
T x^{4}=x_{1}^{4}(t_{1111}+4t_{1112}\frac{x_{2}}{x_{1}}+6t_{1122}(\frac{x_{2}}{x_{1}})^{2}+4t_{1222}(\frac{x_{2}}{x_{1}})^{3}+t_{2222}(\frac{x_{2}}{x_{1}})^{4}).
$$
Obviously, $T x^{4}\geq0$ if and only if
\begin{equation*}
f(s)=\alpha s^{4}+\beta s^{3}+\gamma s^{2}+\mu s+\eta\geq0,
\end{equation*}
where $\alpha=t_{2222},$ $\beta=4t_{1222},$ $\gamma=6t_{1122},$ $\mu=4t_{1112},$ $\eta=t_{1111},$ $s=\displaystyle\frac{x_{2}}{x_{1}}>0$.
Then we have
\begin{align*}
\Delta&=4(12\alpha\eta-3\beta\mu+\gamma^{2})^{3}-(72\alpha\gamma\eta+9\beta\gamma\mu-2\gamma^{3}-27\alpha\mu^{2}-27\beta^{2}\eta)^{2}\\
&=4(12t_{1111}t_{2222}-12\times4t_{1112}t_{1122}+12\times3t_{1222}^{2})^{3}\\
&\quad-(72\times6t_{1111}t_{1122}t_{2222}+72\times12t_{1112}t_{1122}t_{1222}-72\times6t_{1122}^{3}\\
&\quad-72\times6t_{1112}^{2}t_{2222}-72\times6t_{1111}t_{1122}^{2})^{2}\\
&=4\times12^{3}(t_{1111}t_{2222}-4t_{1112}t_{1122}+3t_{1222}^{2})^{3}\\
&\quad-72^{2}\times6^{2}(t_{1111}t_{1122}t_{2222}+2t_{1112}t_{1122}t_{1222}-t_{1122}^{3}-t_{1112}^{2}t_{2222}-t_{1111}t_{1122}^{2})^{2}.
\end{align*}
So the conclusions could be obtained by Lemma \ref{lem24}.
\end{proof}

Now we consider the copositivity of 4th order 3-dimensional symmetric tensor given by scalar dark matter stable
(\cite{BK2012,BK2014,C15,C16,C17}). The most general scalar
potential of the standard model Higgs $H_1$, an inert doublet $H_2$ and a complex singlet $S$ under a $\mathbb{Z}_3$ discrete group is
\begin{equation}\label{H1}\aligned V(H_1,H_2,S)&=\lambda_{1}|H_{1}|^{4}+\lambda_{2}|H_{2}|^{4}+\lambda_{3}|H_{1}|^{2}|H_{2}|^{2}+\lambda_{4}(H_{1}^{\dag}H_{2})(H_{2}^{\dag}H_{1})\\
&\quad+\lambda_{S}|S|^{4}+\lambda_{S1}|S|^{2}|H_{1}|^{2}+\lambda_{S2}|S|^{2}|H_{2}|^{2}\\
&\quad+\frac{1}{2}(\lambda_{S12}S^{2}H_{1}^{\dag}H_{2}+\lambda_{S12}^{*}S^{\dag2}H_{2}^{\dag}H_{1}).
\endaligned\end{equation}
Then the inequality $V(H_1,H_2,S)\geq0$ guarantees the vacuum stability of $\mathbb{Z}_3$ scalar dark matter. By a simple calculation, the inequality $V(H_1,H_2,S)\geq0$ is equivalent to  $R(h_{1},h_{2},s)\geq0,$
\begin{equation} \begin{split} \label{H}
R(h_{1},h_{2},s)&=\lambda_{1}h_{1}^{4}+\lambda_{2}h_{2}^{4}+\lambda_{3}h_{1}^{2}h_{2}^{2}+\lambda_{4}\rho^{2}h_{1}^{2}h_{2}^{2}\\
&\quad+\lambda_{S}s^{4}+\lambda_{S1}s^{2}h_{1}^{2}+\lambda_{S2}s^{2}h_{2}^{2}-|\lambda_{S12}|\rho s^{2}h_{1}h_{2},
\end{split} \end{equation}
where
$ h_{1}=|H_{1}|$, $h_{2}=|H_{2}|$, $H_1^{\dag}H_2=h_{1}h_{2}\rho e^{i\phi}$, $S=se^{i\phi_{S}}$,  $\rho|\in[0,1]$ is the orbit space parameter.
Without loss of generality, we may assume that $h_1^2+h_2^2+s^2=1$ in the sequel.

\begin{theorem}\label{T32} Let $\lambda_1>0$, $\lambda_2>0$, $\lambda_S>0$ and $\rho_0=\frac{|\lambda_{S12}| s^2}{2\lambda_4 h_1h_2}$. Then for all $h_1\geq 0$, $h_2\geq 0$ and $s\geq 0$,
$V(H_1,H_2,S)\geq 0\ (>0)$ if and only if $\lambda_{S2}+2\sqrt{\lambda_2\lambda_S}\geq0\ (>0)$, $\lambda_{S1}+2\sqrt{\lambda_1\lambda_S}\geq0\ (>0)$ and
$$\begin{cases}
   \lambda_3+\lambda_4+2\sqrt{\lambda_1\lambda_2}\geq0\ (>0), &R_{\rho=1}(h_1,h_2,s)\geq0\ (>0), \ \mbox{ if } \lambda_4\leq 0 \\
   \lambda_3+2\sqrt{\lambda_1\lambda_2}\geq0\ (>0),   &R_{\rho=\rho_0,1}(h_1,h_2,s)\geq0\ (>0),  \mbox{ if } \lambda_4> 0.
  \end{cases}$$
\end{theorem}
\begin{proof} It is clear that $R(h_1,0,0)=\lambda_1>0$, $R(0,h_2,0)=\lambda_2>0,$ $R(0,0,s)=\lambda_S>0$ and
$$R(0,h_2,s)=\lambda_2h_2^4+\lambda_Ss^4+\lambda_{S2}h_2^2s^2=\left(
h_2^2 \ \  s^2
\right)\left(
\begin{array}{cc}
\lambda_2 & \frac12\lambda_{S2}\\
\frac12\lambda_{S2} & \lambda_S\\
\end{array}
\right)\left(
\begin{aligned}
&h_2^2\\ &s^2
\end{aligned}
\right).$$
According to Lemma \ref{lem21}, we have \begin{equation}\label{3}R(0,h_2,s)\geq0\ (>0)\mbox{ if and only if }\lambda_{S2}+2\sqrt{\lambda_2\lambda_S}\geq0\ (>0).\end{equation}
Likewise, we also have \begin{equation}\label{4}R(h_1,0,s)\geq0\ (>0)\mbox{ if and only if }\lambda_{S1}+2\sqrt{\lambda_1\lambda_S}\geq0\ (>0),\end{equation}
$$R(h_1,h_2,0)\geq0\ (>0)\mbox{ if and only if }\lambda_3+\lambda_4\rho^2+2\sqrt{\lambda_1\lambda_2}\geq0\ (>0).$$
Since $\rho\in[0,1]$, then the function $f(\rho)=\lambda_3+\lambda_4\rho^2+2\sqrt{\lambda_1\lambda_2}$ reaches its minimum value at $\rho=0$ (if $\lambda_4>0$), $\rho=1$ (if $\lambda_4\leq0$), and hence, \begin{equation}\label{5}\begin{aligned}R(h_1,h_2,0)\geq0\ (>0)\mbox{ if and only if }&\lambda_3+2\sqrt{\lambda_1\lambda_2}\geq0\ (>0)\mbox{ or}\\ &\lambda_3+\lambda_4+2\sqrt{\lambda_1\lambda_2}\geq0\ (>0).\end{aligned} \end{equation}

For $h_1>0$, $h_2>0$ and $s>0$, we consider the function $g(\rho)=R(h_1,h_2,s)$ about one variable $\rho$, which is a quadratic function. Clearly, $$\frac{dg(\rho)}{d\rho}=2\lambda_4\rho h_1^2h_2^2-|\lambda_{S12}| s^2h_1h_2,$$
therefore, the function $g(\rho)$ has a unique extremum value at $\rho_0=\frac{|\lambda_{S12}| s^2}{2\lambda_4 h_1h_2}$.

If $\lambda_4>0$, then $g(\rho)$ reaches its minimum value at $\rho_0=\frac{|\lambda_{S12}| s^2}{2\lambda_4 h_1h_2}\leq1$ or at $\rho=1$ ($\rho_0>1$) and hence, $R(h_1,h_2,s)\geq0$ is now equivalent to $g(\rho_0)\geq0$ or $g(1)\geq0$.
 When $\lambda_4\leq 0$, $g(\rho)$ reaches its minimum value at $\rho=1$, and by that time, $R(h_1,h_2,s)\geq0$ if and only if $g(1)\geq0$. This completes the proof.\end{proof}

Let $x=(x_{1},x_{2},x_{3})^{T}=(h_{1},h_{2},s)^{T}$, and let $\Upsilon=(\tau_{ijkl})$ be a 4th order 3-dimensional symmetric tensor with its entries,
\begin{equation} \label{6}\begin{aligned}
& \tau_{1111}=\lambda_{1},\ \tau_{2222}=\lambda_{2},\ \tau_{3333}=\lambda_{S},\ \tau_{1122}=\displaystyle\frac{1}{6}(\lambda_{3}+\lambda_{4}),\ \tau_{1133}=\displaystyle\frac{1}{6}\lambda_{S1},\\ &  \tau_{2233}=\displaystyle\frac{1}{6}\lambda_{S2},\
 \tau_{1233}=-\displaystyle\frac{1}{12}|\lambda_{S12}|,\  \tau_{ijkl}=0\mbox{ for others.}
\end{aligned}\end{equation}
Then  $g(1)=R_{\rho=1}(h_{1},h_{2},s)=\Upsilon x^{4}$,  and so, the inequality that $R_{\rho=1}(h_{1},h_{2},s)\geq0$ may be transformed into checking the copositivity of $\Upsilon$. By special structure of $\Upsilon$, we now provide the sufficient and necessary conditions of its copositivity. Let \begin{equation} \label{7}\begin{aligned}
\mu_{0}&=4\lambda_{S}\lambda_{1}-\lambda_{S1}^{2},\ \mu_{1}=2\lambda_{S1}|\lambda_{S12}|,\\
\mu_{2}&=4\lambda_{S}\lambda_{3}+4\lambda_{S}\lambda_{4}-|\lambda_{S12}|^{2}-2\lambda_{S1}\lambda_{S2},\\
\mu_{3}&=2\lambda_{S2}|\lambda_{S12}|,\ \mu_{4}=4\lambda_{S}\lambda_{2}-\lambda_{S2}^{2},\\
\Delta&=4(12\mu_{0}\mu_{4}-3\mu_{1}\mu_{3}+\mu_{2})^{3}\\
&-(72\mu_{0}\mu_{2}\mu_{4}+9\mu_{1}\mu_{2}\mu_{3}-2\mu_{2}^3-27\mu_{0}
\mu_{3}^2-27\mu_{1}^2\mu_{4})^{2}.
\end{aligned}\end{equation}

\begin{theorem}\label{T33} Let $\Upsilon=(\tau_{ijkl})$ given by (\ref{6}) with $\lambda_1>0$, $\lambda_2>0$, $\lambda_S>0$. Then
$\Upsilon$ is copositive if and only if
\begin{flushleft}
\quad(1) $\lambda_{S1}\geq0,$ $\lambda_{S2}\geq0,$ $2\sqrt{\lambda_{S1}\lambda_{S2}}\geq|\lambda_{S12}|$,
\  $\lambda_{3}+\lambda_{4}+2\sqrt{\lambda_{1}\lambda_{2}}\geq0$.
\end{flushleft}
\begin{flushleft}
\quad(2) $\lambda_{S1}<0,$ $\lambda_{S2}<0$, $4\lambda_{S}\lambda_{2}-\lambda_{S2}^{2}>0,$ $4\lambda_{S}\lambda_{1}-\lambda_{S1}^{2}>0$ and
\end{flushleft}
\begin{flushleft}
\qquad\textcircled{1}$\Delta\leq0$, $\mu_{3}\sqrt{\mu_{0}}+\mu_{1}\sqrt{\mu_{4}}>0$, or
\end{flushleft}
\begin{flushleft}
\qquad\textcircled{2}$\Delta\geq0$, $|\mu_{1}\sqrt{\mu_{4}}-\mu_{3}\sqrt{\mu_{0}}|\leq4\sqrt{\mu_{0}\mu_{2}\mu_{4}+2\mu_{0}\mu_{4}\sqrt{\mu_{0}\mu_{4}}}$,\\
\end{flushleft}
\begin{flushleft}
\quad\qquad(i) $-2\sqrt{\mu_{0}\mu_{4}}\leq\mu_{2}\leq6\sqrt{\mu_{0}\mu_{4}}$;
\end{flushleft}
\begin{flushleft}
\quad\qquad(ii) $\mu_{2}>6\sqrt{\mu_{0}\mu_{4}}$,
\end{flushleft}
\begin{flushleft}
\qquad\qquad$\mu_{1}\sqrt{\mu_{4}}+\mu_{3}\sqrt{\mu_{0}}\geq-4\sqrt{\mu_{0}\mu_{2}\mu_{4}-2\mu_{0}\mu_{4}
\sqrt{\mu_{0}\mu_{4}}}$.
\end{flushleft}
\end{theorem}

\begin{proof} Rewritten the equation (\ref{H}) as follows, \begin{equation*}
\Upsilon x^{4}=V(h_1,h_2,s)=\lambda_{S}t^2+\Phi(h_{1},h_{2})t+\widetilde{\Upsilon}(h_{1},h_{2}),
\end{equation*}where $t=s^{2}$, \begin{equation} \label{M}
 \Phi(h_{1},h_{2})=\lambda_{S1}h_{1}^{2}+\lambda_{S2}h_{2}^{2}-|\lambda_{S12}| h_{1}h_{2},
\end{equation}
\begin{equation} \label{V}
 \widetilde{\Upsilon}(h_{1},h_{2})=\lambda_{1}h_{1}^{4}+\lambda_{2}h_{2}^{4}+(\lambda_{3}+\lambda_{4})h_{1}^{2}h_{2}^{2}.
\end{equation}
Then $\Upsilon x^{4}$ may be seen as a  quadratic polynomial about one variable $t$, and hence, it easily obtained by Lemma \ref{lem23} that $\Upsilon$ is  copositive if and only if
\begin{flushleft}
\quad(1)\  $\Phi(h_{1},h_{2})\geq0,$  $\widetilde{\Upsilon}(h_{1},h_{2})\geq0$,
\end{flushleft}
\begin{flushleft}
\quad(2)\  $\Phi(h_{1},h_{2})<0,$  $4\lambda_{S}\widetilde{\Upsilon}(h_{1},h_{2})-(\Phi(h_{1},h_{2}))^{2}\geq0.$
\end{flushleft}

 Obviously, both $\Phi(h_{1},h_{2})$ and $\widetilde{\Upsilon}(h_{1},h_{2})$ are quadratic forms with coefficient matrices
$$\left(
\begin{array}{cc}
\lambda_{S1} & -\frac{1}{2}|\lambda_{S12}|\\
-\frac{1}{2}|\lambda_{S12}| & \lambda_{S2}\\
\end{array}\right)\mbox{ and }\left(
\begin{array}{cc}
\lambda_{1} & \frac{1}{2}(\lambda_{3}+\lambda_{4})\\
\frac{1}{2}(\lambda_{3}+\lambda_{4}) & \lambda_{2}\\
\end{array}
\right),$$
and so, it follows from Lemma \ref{lem21} that
$\Phi(h_{1},h_{2})\geq0$ and $\widetilde{\Upsilon}(h_{1},h_{2})\geq0$ are respectively equivalent to $$
\lambda_{S1}\geq0,\ \lambda_{S2}\geq0, \ -|\lambda_{S12}|+2\sqrt{\lambda_{S1}\lambda_{S2}}\geq0\mbox{ and }
\lambda_{3}+\lambda_{4}+2\sqrt{\lambda_{1}\lambda_{2}}\geq0.$$

Similarly, the inequality that $\Phi(h_{1},h_{2})<0$ is equivalent to the strict copositivity of $-\Phi(h_{1},h_{2})$, and then,  $\Phi(h_{1},h_{2})<0$ if and only if
$$\lambda_{S1}<0, \ \lambda_{S2}<0,\ |\lambda_{S12}|+2\sqrt{\lambda_{S1}\lambda_{S2}}>0.$$
It is always tenable that $|\lambda_{S12}|+2\sqrt{\lambda_{S1}\lambda_{S2}}>0$, and hence, $\Phi(h_{1},h_{2})<0$ if and only if
$$\lambda_{S1}<0, \ \lambda_{S2}<0.$$

Now we prove that $4\lambda_{S}\widetilde{\Upsilon}(h_{1},h_{2})-(\Phi(h_{1},h_{2}))^{2}\geq0$. We may rewrite this equation as follows,
\begin{align*}
4\lambda_{S}\widetilde{\Upsilon}(h_{1},h_{2})-(\Phi(h_{1},h_{2}))^{2}
&=4\lambda_{S}(\lambda_{1}h_{1}^{4}+\lambda_{2}h_{2}^{4}+(\lambda_{3}+\lambda_{4})h_{1}^{2}h_{2}^{2})\\
&\quad-(\lambda_{S1}h_{1}^{2}+\lambda_{S2}h_{2}^{2}-|\lambda_{S12}| h_{1}h_{2})^{2}\\
&=(4\lambda_{S}\lambda_{1}-\lambda_{S1}^{2})h_{1}^{4}+2\lambda_{S1}|\lambda_{S12}| h_{1}^{3}h_{2}\\
&\quad+(4\lambda_{S}\lambda_{3}+4\lambda_{S}\lambda_{4}-|\lambda_{S12}|^{2}-2\lambda_{S1}\lambda_{S2})h_{1}^{2}h_{2}^{2}\\
&\quad+2\lambda_{S2}|\lambda_{S12}| h_{1}h_{2}^{3}+(4\lambda_{S}\lambda_{2}-\lambda_{S2}^{2})h_{2}^{4}\\
&=\mu_{0}h_{1}^{4}+\mu_{1}h_{1}^{3}h_{2}+\mu_{2}h_{1}^{2}h_{2}^{2}+\mu_{3}h_{1}h_{2}^{3}+\mu_{4}h_{2}^{4}.
\end{align*}
Naturely, this defines a 4th order 2-dimensional symmetric tensor $T=(t_{ijkl})$ with its entries
\begin{equation*}
t_{1111}=\mu_{0},\ t_{2222}=\mu_{4},\ t_{1112}=\frac{1}{4}\mu_{1},\ t_{1122}=\frac{1}{6}\mu_{2}, \ t_{1222}=\frac{1}{4}\mu_{3}.
\end{equation*}
Then by Theorem \ref{T31}, we have \begin{flushleft}
\qquad\textcircled{1}$\Delta\leq0$, $\mu_{3}\sqrt{\mu_{0}}+\mu_{1}\sqrt{\mu_{4}}>0$, or
\end{flushleft}
\begin{flushleft}
\qquad\textcircled{2} $\mu_{3}\geq0$, $\mu_{1}\geq0$, $\mu_{2}+2\sqrt{\mu_{0}\mu_{4}}\geq0$, or
\end{flushleft}
\begin{flushleft}
\qquad\textcircled{3}$\Delta\geq0$, $|\mu_{1}\sqrt{\mu_{4}}-\mu_{3}\sqrt{\mu_{0}}|\leq4\sqrt{\mu_{0}\mu_{2}\mu_{4}+2\mu_{0}\mu_{4}\sqrt{\mu_{0}\mu_{4}}}$,\\
\end{flushleft}
\begin{flushleft}
\quad\qquad(i)$-2\sqrt{\mu_{0}\mu_{4}}\leq\mu_{2}\leq6\sqrt{\mu_{0}\mu_{4}}$;
\end{flushleft}
\begin{flushleft}
\quad\qquad(ii)$\mu_{2}>6\sqrt{\mu_{0}\mu_{4}}$,
\end{flushleft}
\begin{flushleft}
\qquad\qquad$\mu_{1}\sqrt{\mu_{4}}+\mu_{3}\sqrt{\mu_{0}}\geq-4\sqrt{\mu_{0}\mu_{2}\mu_{4}-2\mu_{0}\mu_{4}
\sqrt{\mu_{0}\mu_{4}}}$.
\end{flushleft}
Because of  $\lambda_{S1}<0$ and $\lambda_{S2}<0$,  there will be no the inequalities $\mu_{3}=2\lambda_{S1}|\lambda_{S12}|\geq0$ and $\mu_{1}=2\lambda_{S2}|\lambda_{S12}|\geq0$, and hence, the above conditions $\textcircled{1}$ and $\textcircled{3}$ guarantee that  $4\lambda_{S}\widetilde{\Upsilon}(h_{1},h_{2})-(\Phi(h_{1},h_{2}))^{2}\geq0$.  This completes the proof.\end{proof}

Next we show the necessary and sufficient conditions of the inequality $R_{\rho=\rho_0}(h_1,h_2,s)=g(\rho_0)\geq0$.

\begin{theorem} \label{T34} Let $\lambda_1>0$, $\lambda_2>0$, $\lambda_S>0$, $\lambda_4>0$ and $\rho_0=\frac{|\lambda_{S12}| s^2}{2\lambda_4 h_1h_2}$. Then
  $R_{\rho=\rho_0}(h_{1},h_{2},s)\geq0$ if and only if
  \begin{flushleft}
\qquad $4\lambda_4\lambda_{S}-|\lambda_{S12}|^2\geq0$, $\alpha=\lambda_3+2\sqrt{\lambda_1\lambda_2}\geq0$,
\end{flushleft}
\begin{flushleft}
\qquad$\beta=\lambda_{S1}+2\sqrt{\lambda_1(\lambda_{S}-\frac{|\lambda_{S12}|^2}{4\lambda_4}})\geq0$, $\gamma=\lambda_{S2}+2\sqrt{\lambda_2(\lambda_{S}-\frac{|\lambda_{S12}|^2}{4\lambda_4}})\geq0$,
\end{flushleft}
\begin{flushleft}
\qquad $\lambda_3\sqrt{\lambda_{S}-\frac{|\lambda_{S12}|^2}{4\lambda_4}}+\lambda_{S1}\sqrt{\lambda_2}+\lambda_{S2}\sqrt{\lambda_1}+\sqrt{\alpha\beta\gamma}
\geq0$.
\end{flushleft}
\end{theorem}
\begin{proof} We plug $\rho_0=\dfrac{|\lambda_{S12}| s^2}{2\lambda_4 h_1h_2}$ into the equation \eqref{H},
  \begin{align}
    R_{\rho=\rho_0}(h_{1},h_{2},s) & =\lambda_1h_{1}^{4}+\lambda_{2}h_{2}^{4}+\lambda_{S}s^{4}+\frac{|\lambda_{S12}|^2}{4\lambda_4}s^{4}-\frac{|\lambda_{S12}|^2}{2\lambda_4}s^{4}\nonumber\\
   &\quad+\lambda_{3}h_{1}^{2}h_{2}^{2}+\lambda_{S1}s^{2}h_{1}^{2}+\lambda_{S2}s^{2}h_{2}^{2}\nonumber\\
   &= \lambda_1h_{1}^{4}+\lambda_{2}h_{2}^{4}+(\lambda_{S}-\frac{|\lambda_{S12}|^2}{4\lambda_4})s^{4}\nonumber\\
   &\quad+\lambda_{3}h_{1}^{2}h_{2}^{2}+\lambda_{S1}s^{2}h_{1}^{2}+\lambda_{S2}s^{2}h_{2}^{2}.\label{10}
  \end{align}
Then $R_{\rho=\rho_0}(h_{1},h_{2},s)$ may be seen as a quadratic form about $(h_{1}^2,h_{2}^2,s^2)$ with the coefficient matrix
\begin{equation}\label{11}\left(
\begin{array}{ccc}
\lambda_1 & \frac{1}{2}\lambda_3&\frac{1}{2}\lambda_{S1}\\
\frac{1}{2}\lambda_3 & \lambda_2&\frac{1}{2}\lambda_{S2}\\
\frac{1}{2}\lambda_{S1}&\frac{1}{2}\lambda_{S2} &\lambda_{S}-\frac{|\lambda_{S12}|^2}{4\lambda_4}\\
\end{array}\right).\end{equation}
Therefore, $R_{\rho=\rho_0}(h_{1},h_{2},s)\geq0$ can be transformed into checking  copositivity of the above coefficient matrix, and hence, after making simple calculations, the desired results immediately obtain from Lemma \ref{lem21}.
\end{proof}

In summary, we prove that $R_{\rho=1}(h_{1},h_{2},s)\geq0$  and $R_{\rho=\rho_0}(h_{1},h_{2},s)\geq0$ in Theorems \ref{T33} and \ref{T34}, and then, by Theorem \ref{T32}, the following corollary is obtained easily.

\begin{corollary}\label{T5} Let $V(H_1,H_2,S)$ be given by \eqref{H1} with $\lambda_1>0$, $\lambda_2>0$, $\lambda_S>0$. Then $V(H_1,H_2,S)\geq 0$ if and only if  $\lambda_{S2}+2\sqrt{\lambda_2\lambda_S}\geq0$, $\lambda_{S1}+2\sqrt{\lambda_1\lambda_S}\geq0$ and
\begin{flushleft}
\quad (I) $\lambda_{4}\leq0$, $\lambda_{3}+\lambda_{4}+2\sqrt{\lambda_{1}\lambda_{2}}\geq0$ and
\end{flushleft}
\begin{flushleft}
\qquad(1) $\lambda_{S1}\geq0,$ $\lambda_{S2}\geq0,$ $2\sqrt{\lambda_{S1}\lambda_{S2}}\geq|\lambda_{S12}|$.
\end{flushleft}
\begin{flushleft}
\qquad(2) $\lambda_{S1}<0,$ $\lambda_{S2}<0$, $4\lambda_{S}\lambda_{2}-\lambda_{S2}^{2}>0,$ $4\lambda_{S}\lambda_{1}-\lambda_{S1}^{2}>0$ and
\end{flushleft}
\begin{flushleft}
\quad\qquad\textcircled{1}$\Delta\leq0$, $\mu_{3}\sqrt{\mu_{0}}+\mu_{1}\sqrt{\mu_{4}}>0$,
\end{flushleft}
\begin{flushleft}
\quad\qquad\textcircled{2}$\Delta\geq0$, $|\mu_{1}\sqrt{\mu_{4}}-\mu_{3}\sqrt{\mu_{0}}|\leq4\sqrt{\mu_{0}\mu_{2}\mu_{4}+2\mu_{0}\mu_{4}\sqrt{\mu_{0}\mu_{4}}}$,\\
\end{flushleft}
\begin{flushleft}
\qquad\qquad(i)$-2\sqrt{\mu_{0}\mu_{4}}\leq\mu_{2}\leq6\sqrt{\mu_{0}\mu_{4}}$;
\end{flushleft}

\begin{flushleft}
\qquad\qquad(ii)$\mu_{2}>6\sqrt{\mu_{0}\mu_{4}}$,
\end{flushleft}
\begin{flushleft}
\quad\qquad\qquad$\mu_{1}\sqrt{\mu_{4}}+\mu_{3}\sqrt{\mu_{0}}\geq-4\sqrt{\mu_{0}\mu_{2}\mu_{4}-2\mu_{0}\mu_{4}
\sqrt{\mu_{0}\mu_{4}}}$.
\end{flushleft}
\begin{flushleft}
\quad (II) $\lambda_{4}>0$,  $\alpha=\lambda_3+2\sqrt{\lambda_1\lambda_2}\geq0$, either $4\lambda_4\lambda_{S}-|\lambda_{S12}|^2\geq0$,
\end{flushleft}
\begin{flushleft}
\qquad$\beta=\lambda_{S1}+2\sqrt{\lambda_1(\lambda_{S}-\frac{|\lambda_{S12}|^2}{4\lambda_4}})\geq0$, $\gamma=\lambda_{S2}+2\sqrt{\lambda_2(\lambda_{S}-\frac{|\lambda_{S12}|^2}{4\lambda_4}})\geq0$,
\end{flushleft}
\begin{flushleft}
\qquad $\lambda_3\sqrt{\lambda_{S}-\frac{|\lambda_{S12}|^2}{4\lambda_4}}+\lambda_{S1}\sqrt{\lambda_2}+\lambda_{S2}\sqrt{\lambda_1}+\sqrt{\alpha\beta\gamma}
\geq0$;
\end{flushleft}
\begin{flushleft}
\qquad or (1) and (2) of (I) hold,
\end{flushleft}
where $\Delta$ and $\mu_i$ ($i=0,1,2,3,4$) are given by \eqref{7}.
\end{corollary}

\begin{remark}
The analytic necessary and sufficient conditions of copositivity for a special tensor $\Upsilon(\rho)$ with a parameter $\rho$ is proved, where $\Upsilon(\rho)=(\tau_{ijkl})$ is a 4th order 3-dimensional symmetric real tensor with its entries, \begin{flushleft}
\qquad$\tau_{1111}=\lambda_{1},$ $\tau_{2222}=\lambda_{2},$ $\tau_{3333}=\lambda_{S}$, $\tau_{1122}=\displaystyle\frac{1}{6}(\lambda_{3}+\lambda_{4}\rho^{2}),$ $\tau_{1133}=\displaystyle\frac{1}{6}\lambda_{S1},$
\end{flushleft}
\begin{flushleft}
\qquad $\tau_{2233}=\displaystyle\frac{1}{6}\lambda_{S2}$, $\tau_{1233}=-\displaystyle\frac{1}{12}|\lambda_{S12}|\rho,$ $\tau_{ijkl}=0$ for others.
\end{flushleft}
 However, the analytic expressions of strict copositivity of such a tensor is still unknown.  Then for a general 4th order 3-dimension symmetric real tensor, how to obtain its analytic expressions of (strict) copositivity, which is worthy to be studied further.
\end{remark}
\begin{remark} For three Higgs doublets with equal electrically weak quantum numbers $\mu_i,\ i = 1, 2, 3$, the
Higgs potential model can be constructed as the following form \cite{DIK,I2020,IV2012,IV2013,MR,IK},
\begin{align*}
  R= & -\frac{M_0}{\sqrt{3}}(\mu_1^*\mu_1+\mu_2^*\mu_2+\mu_3^*\mu_3)+\frac{\Lambda_0}{3}(\mu_1^*\mu_1+\mu_2^*\mu_2+\mu_3^*\mu_3)^2\\
   & +\frac{\Lambda_3}{3}\left[(\mu_1^*\mu_1)^2+(\mu_2^*\mu_2)^2+(\mu_3^*\mu_3)^2-(\mu_1^*\mu_1)(\mu_2^*\mu_2)-(\mu_1^*\mu_1)(\mu_3^*\mu_3)-(\mu_2^*\mu_2)(\mu_3^*\mu_3) \right]\\
   & +\Lambda_1\left[(\mbox{Re}\mu_1^*\mu_2)^2+(\mbox{Re}\mu_2^*\mu_3)^2+(\mbox{Re}\mu_3^*\mu_1)^2\right] \\
   & +\Lambda_2\left[(\mbox{Im}\mu_1^*\mu_2)^2+(\mbox{Im}\mu_2^*\mu_3)^2+(\mbox{Im}\mu_3^*\mu_1)^2\right] \\
   & +\Lambda_4\left[(\mbox{Re}\mu_1^*\mu_2)(\mbox{Im}\mu_1^*\mu_2)+(\mbox{Re}\mu_2^*\mu_3)(\mbox{Im}\mu_2^*\mu_3)+(\mbox{Re}\mu_3^*\mu_1)(\mbox{Im}\mu_3^*\mu_1)\right].
\end{align*}
Then how to solve the analytic necessary and sufficient conditions of the boundedness from below of the above model ($R\geq0$) is a topic worthy of study and practical significance. It may be seen as a 4th order 3-dimensional symmetric tensor on complex field, and so, this problem is converted into a problem of positive definiteness (or copositivity) of the corresponding  tensor.
\end{remark}

\begin{remark}
It is known from Song-Qi \cite{C31} that  the (strict) copositivity of a symmetric tensor $\mathcal{T}=(t_{i_1\cdots i_m})$ is equivalent to the uniqueness of solution to the tensor
complementary problem TCP($\mathcal{T},q$) for $q>0$ ($q\geq0$),  $$x\geq0,\ \mathcal{T} x^{m-1}+q\geq0,\ x^\top(\mathcal{T} x^{m-1}+q)=0.$$ So, to test the strict copositivity of such a special tensor $\Upsilon(\rho)$, an alternative method may be to solve the  TCP($\Upsilon(\rho),q$), $$x\geq0,\ \Upsilon(\rho)x^3+q\geq0,\ x^\top(\Upsilon(\rho)x^3+ q)=0.$$
Then how to obtain a solution $x$ of  TCP($\Upsilon(\rho),q$) with a parameter $\rho$, which deserves further research.
\end{remark}


\section{Conclusions}

In this article,  we discuss the analytic necessary and sufficient conditions of copositivity for a class of special symmetric tensors given by vacuum stability model.  This actually presents how to identify copositivity of a 4th order tensor with a parameter. Moreover, this work first provides a way to test the uniqueness of solution to tensor complementary problem with a parameter.

\begin{acknowledgements} The authors would like to express their sincere thanks to the editors and anonymous referees for his/her constructive comments and valuable suggestions.
 This  work was supported by
			the National Natural Science Foundation of P.R. China (Grant No.
			12171064), by The team project of innovation leading talent in Chongqing (No.CQYC20210309536) and by the Foundation of Chongqing Normal University (20XLB009)
\end{acknowledgements}

\end{document}